# Towards Approaches to Continuous Assessment of Cyber Risk in Security of Computer Networks

Alexander Kott and Curtis Arnold
US Army Research Laboratory, Adelphi, MD

From mobile devices to cloud computing, Government technology officials have high hopes for benefits of continuous monitoring and risk scoring (CMRS). Indeed, benefits could be great, but so are the challenges of implementing a successful CMRS system.

For authoritative definitions related to CMRS one may consult NIST publication SP 800-137 [NIST 2011], where Information Security Continuous Monitoring is defined as "… maintaining ongoing awareness of information security, vulnerabilities, and threats to support organizational risk management decisions, " or DISA, which defines CMRS as "…visibility of cyber risks and demonstrates the ability to use DOD Enterprise security tools and content," and continuous monitoring as "…the on–going observation, assessment, analysis, and diagnosis of an organization's cybersecurity posture, hygiene, and operational readiness" (http://www.disa.mil/Services/Information-Assurance/SCM/CM-Definition )

For the purposes of this brief article, we focus on two most salient aspects of CMRS. First, continuous collection of data through automated feeds; hence the "continuous monitoring." Typical data collected for continuous monitoring purposes include network traffic information as well as host information from host-based agents: vulnerability information and patch status about hosts on the network; scan results from tools like Nessus; TCP netflow data; DNS trees, etc. These diverse types of information must be collected with appropriate speed, translated from multiple formats, correlated, fused, organized and stored for further processing.

Second, analysis of the collected data in order to assess the risks – the "risk scoring." This assessment may include flagging especially egregious vulnerabilities and exposures, or computing metrics that provide an overall characterization of the network's risk level. Currently used risk metrics are often simple sums or counts of vulnerabilities and missing patches. More on this shortly.

The first and best known example of a successful CMRS system is the iPost developed by the Department of State [iPost 2010]. In 2009, iPost won NSA Frank B. Rowlett Award. It became a basis for follow-on efforts to extend and apply similar approaches in other organizations.

Although it is not yet easy to point to major CMRS successes other than iPost, the concept of CMRS attracted strong interest and commitment, particularly in the U.S. Government. Consider these two important examples. Department of Defense has recently released a document on its mobile device strategy [DoD 2012a] that specifies continuous monitoring as part of the device management service. Shortly after that, DoD released a document on cloud computing strategy [DoD 2012b] that also gives a major role to CMRS. In particular, cloud providers hosting DoD data off site will have to integrate their continuous monitoring and response capabilities with U.S. Cyber Command's systems for protecting DoD information.

Such strategic documents can have far-reaching ramifications. Numerous corporations interested in providing or supporting mobile device services and cloud computing services to the U.S. Government will rash to implement a CMRS capability. Same or related practices will spill over to commercial, non-Government sectors.

What causes this explosion of interest in CMRS? There are multiple forcing functions, often of financial nature. For example, the Federal Information Security Management Act of 2002 (FISMA) imposes on government agencies rigorous, annual reporting requirements. DoD programs like CCRI and DIACAP impose a three-year cycle of Certification and Accreditation requirements.

Such requirements were quickly found to be exceedingly expensive: labor intensive and a major management burden. Furthermore, a year or three years are such astronomically long times in our age of daily technological changes that the accreditation process is hardly effective in terms of enhanced security. To have any degree of effectiveness, such an accreditation should be performed far more often, but who can afford this when even a three-year cycle is prohibitively costly? Enter CMRS. It promises a potential reduction in costs along with dramatic improvements in timeliness and reliability of risk awareness and remedies.

The reduction of costs and risks is comprised of several key components. Importantly, CMRS is expected to reduce costs associated with regulatory compliance and accreditation. Automated data feeds – many of which are already available within enterprise system and can be used for CMRS purposes – along with semi-automated risk assessments, can reduce costs associated with conventional data gathering for accreditation. Management burden is likely to diminish accordingly. Because CMRS provides near continuous visibility of assets, costly and dangerous situations such as rogue devices, mis-configurations and failures of patching are recognized and corrected rapidly and with less need for multiple, costly security controls. If a security control diminishes in its effectiveness, continuous monitoring is likely to highlight the trend and encourage timely changes.

Particularly promising are the benefits of automated quantification of risk, i.e., of assigning risk scores or other numerical measures to the network as w hole, its subsets and even individual assets. This opens doors to true risk management decision-making [Kott 2006; Kott 2014], potentially highly rigorous and insightful. Employees at multiple levels – from senior leaders to system administrators – will be aware of continually updated risk distribution over the network components, and will use this awareness to prioritize application of resources to most effective remedial actions. Quantification of risks can also contribute to rapid, automated or semi-automated implementation of remediation plans.

Great benefits, however, come with major challenges, and CMRS is no exception. These challenges fall mainly into two categories. The first centers on the problem of integrating and fusing highly heterogeneous information. Continuous monitoring implies consistent, largely centralized collection and processing of information – a difficult objective to achieve in any large-scale, geographically distributed and functionally diverse enterprise system. Developers of CMRS systems must deal with broad heterogeneity of technologies on the network, including that of the technologies that obtain and provide data feeds to the CMRS data collection mechanisms. These data – very diverse in their nature and content -- arrive in widely differing formats, often with uncertain and imprecise semantics, with timeliness, availability and completeness varying between the technologies and across different segments of the network. Departments and agencies have different missions, constraints, business process and corresponding differences in systems. Hopes for increasing uniformity and consistent standardization are unlikely to be fulfilled; heterogeneity of our systems will continue to increase.

Data integration benefits from availability and continuing development of standards like the Security Content Automation Program (SCAP) that standardizes identification and description of security issues and includes Common Vulnerabilities and Exposures (CVE) dictionary, the National Vulnerability Database (NVD) that relates vulnerabilities and software, and the Common Vulnerability Scoring System (CVSS) that provides a severity score for each vulnerability. SCAP has been a major enabler of CMRS. However, even if integration of data is achieved successfully, a larger challenge still looms – the challenge of information fusion, a technically profound problem that has been a subject of extensive research for decades, and will likely remain such for many years to come.

Part of dealing with heterogeneity of data and systems is to establish an overarching strategy and architecture for integration of these widely different data flows. This requires an agreement among potentially a large number of stakeholders, which in turn requires lengthy diplomacy while the high pace of technology threatens to make any major architecture outdated by the time an agreement on an all encompassing architecture is reached. Alternative approach might be a relatively loose confederation of local solutions. Cybersecurity community has recognized the importance of architectural approaches in CMRS and a number of related efforts are underway. For example, DHS collaborates with a number of other departments of the Federal Government in developing and extending the Continuous Asset Evaluation, Situational Awareness and Risk Scoring (CAESARS) reference architecture for continuous monitoring, http://www.dhs.gov/continuous-asset-evaluation-situational-awareness-and-risk-scoring-reference-architecture-report.

The second group of challenges is the lack of rigorous approaches to computing risk. Existing risk scoring algorithms remain limited to ad hoc heuristics such as simple sums of vulnerability scores or counts of things like missing patches or open ports, etc. Weaknesses and potentially misleading nature of such metrics have been pointed out by a number of specialists, e.g., [Jansen 2009; Brennan 2009; Bartol 2009]. For example, the individual vulnerability scores are dangerously reliant on subjective, human, qualitative input, potentially inaccurate and expensive to obtain. Further, the total number of vulnerabilities may matters far less than how vulnerabilities are distributed over hosts, or over time. Similarly, neither topology of the network nor the roles and dynamics of inter-host interactions are considered by simple sums of vulnerabilities or missing patches.

There are numerous others factors that strongly impact risks to a network, and yet are unaccounted for by simple existing metrics. These include: nonlinear interactions between vulnerabilities; importance of assessing risks with respect to adversary intent and capabilities; impossibility of considering risks without an estimate of how a compromise might affect the mission of the organization; need to account for presence or absence of appropriate network defense mechanisms [Kott 2014]. Regarding the particularly serious risks from Advanced Persistent Threats, it is doubtful that any risk scoring based on known vulnerabilities is applicable. In general, there is a pronounced lack of rigorous theory and models of how various factors might combine into quantitative characterization of true risks. There are initial efforts, such as [Lippman 2012] to formulate scientifically rigorous methods of calculating risks, and these should be given strong consideration in practical implementations of CMRS.

Also lacking is rigorous empirical validation of risk metrics, whether they are heuristic or theoretically-grounded. There is little documented empirical evidence that currently used risk scores exhibit correlation with actual risks experienced by systems to which the scores are applied. Although one might find it intuitively compelling that a heuristic metric like a count of major vulnerabilities is a reasonable reflection of true risks, the intuition is no substitute to rigorous evidence, and may be misleading at least in some cases. Implementers and operators of CMRS should not overlook an opportunity to collect empirical data that would remedy this shortage of evidence.

Although the challenges of CMRS are real and significant they should not deter organizations from implementing a CMRS solution. Benefits of CMRS are undoubtedly great and even a partial solution will bring a significant fraction of the benefits. As the collective experience with CMRS solutions grows, the science and technology community will produce new solutions to some of the technical challenges we mentioned here, while others may turn out to be less critical in practice than they appear now.